\documentstyle[epsfig]{l-aa}

\begin{document}

\def\ltsima{$\; \buildrel < \over \sim \;$}
\def\simlt{\lower.5ex\hbox{\ltsima}}

%\large
   \thesaurus{08.02.1 -- 08.14.2 -- 13.25.5 -- 08.09.2}
   \title{The X--ray emission of the polar BL Hydri}

   \author{G. Matt 
          \inst{1}
%   \and the Beppopolars team
    \and R. Barcaroli
          \inst{1}
    \and T. Belloni
          \inst{2}
    \and K. Beuermann
          \inst{3}
    \and J.M. Bonnet--Bidaud 
	 \inst{4}
    \and D. De Martino
          \inst{5}
    \and C. Done
          \inst{6}
    \and B.T. G\"ansicke
          \inst{3}
    \and M. Guainazzi
          \inst{7}
    \and M. Mouchet
          \inst{8,9}
    \and K. Mukai
          \inst{10}
%          \inst{1,} \inst{4}
          }

%   \offprints{G. Matt, matt@amaldi.fis.uniroma3.it}
% \\
%              proposed choice: Main Journal \\
%              proposed section: Formation and evolution of stars \\
%              proofs to: G. Matt, tel.~+39--6--55177024
%fax~+39--6--5579303,
%                         email: matt@amaldi.fis.uniroma3.it}

   \institute{Dipartimento di Fisica, Universit\`a degli studi ``Roma Tre", 
              Via della Vasca Navale 84, I--00146 Roma, Italy
    \and Astronomical Institute ``Anton Pannekoek", University of Amsterdam
		and Center for High-Energy Astrophysics, 
                Kruislaan 403, NL-1098 SJ Amsterdam, The Netherlands
   \and Universit\"atssternwarte G\"ottingen, Geismarlandstrasse 11, D-37083
                 G\"ottingen, Germany
   \and CEA, DSN/DAPNIA/Service d'Astrophysique, CEN Saclay, F-91191 Gif sur
         Yvette Cedex, France
    \and Osservatorio Astronomico di Capodimonte, Via Moiariello 16, 
         	I-80131 Napoli, Italy
    \and Department of Physics, University of Durham, South Road, Durham
                DH1 3LE, U.K.
    \and SAX/SDC Nuova Telespazio, Via Corcolle 19,  I--00131 Roma, Italy
    \and DAEC, Observatoire de Paris, Section de Meudon, F-92195 Meudon Cedex, France
    \and Universit\'e Denis Diderot, 2 Place Jussieu, F-75005 Paris, France
    \and NASA/GSFC, Code 668, Greenbelt MD 20771, USA
   }

   \date{Received / Accepted }

   \maketitle

   \begin{abstract}
	We report on the analysis of the
        ASCA and BeppoSAX X--ray observations of the polar system BL Hyi,
        performed in October 94 and September 96, respectively.

	Emission from both poles is apparent from the folded light curves
        of both observations;
        the emission from the second pole varies from cycle to cycle, 
	indicating non--stationary accretion there. 

	The temperature of the post--shock region is estimated
        to be about 10 keV. Inclusion of both complex absorption
        and Compton reflection significantly improves the quality of the fit. 
        No soft X--ray component is observed; the BeppoSAX/LECS 
	upper limit to the soft component is 
	in agreement with theoretical expectations for this low magnetic field
        system. 
   \end{abstract}

%
%  14.Sep.'90: Demo-Vs.
%________________________________________________________________

\section{Introduction}

 Polar systems, a subgroup of magnetic Cataclysmic Variables (mCVs),
 contain a highly magnetized white dwarf with polar field strengths
ranging from  $\sim$ 10
MG $\sim$ to 230 MG (see Beuermann 1997 and references therein), and 
 accreting material from a late type main sequence star. The
magnetic field of the white dwarf is strong enough to phase--lock its
rotation with the orbital period.
These systems are strong X-ray emitters in both soft and hard X-ray bands
 (see review by Cropper 1990).  While hard X-rays are emitted from
a standing shock above the white dwarf surface, 
soft X-rays originate from hot photospheric regions heated
either by irradiation from the post-shock plasma (Lamb \& Masters 1979) or by
dense plasma blobs carrying their kinetic energy deep into the atmosphere
(Kuipers \& Pringle 1982). 
Irradiation is important primarily for flow rates sufficiently low for
the shock to stand high above the surface. However, a large fraction
of the reprocessed radiation appears in the UV rather than at soft
X-rays, as shown quantitatively in the case of AM\,Her 
(G\"ansicke et al. 1995). The ``blobby'' accretion mode is increasingly
important at higher field strengths, because the blobs become more and more
compressed when arriving at the surface of the white dwarf (e.g. Beuermann \&
Burwitz 1995; King 1995 and references therein).

While the soft X--ray component is in general adequately fitted by a
black--body spectrum with a temperature of a few tens of eV (even if more
sophisticated models are sometimes needed: see e.g. Van Teeseling et al. 1994), 
it has recently become clear that a simple thermal plasma
model is no longer adequate in reproducing the hard X--ray component of Polars. 
Reflection
from the white dwarf surface, complex absorption and multi--temperature
emission may contribute significantly to the spectrum above a few tenth of
keV (e.g. Cropper et al. 1997 and references therein).

\begin{table*}
\centering
\caption{ASCA and BeppoSAX observations log}
\label{log}
\vspace{0.05in}
\begin{tabular}{cccccc}
\hline
\hline
~ & ~ & ~ & ~ & ~ & ~\cr
Instrument & En. range & Date of obs. & Exp. 
Time$^a$ & Count rate & 2-10 keV flux$^b$ \cr
~ & (keV)  & ~ & (ks) & (cts/s)  & (erg/s/cm$^2$) \cr
\noalign {\hrule}
~ & ~ & ~ & ~ & ~ & ~\cr
ASCA/GIS2 & 0.8-10 & 1994 Oct 11-12 & 43 & 0.13 & 8.0$\times$10$^{-12}$ \cr
ASCA/GIS3 & 0.8-10 & 1994 Oct 11-12 & 43 & 0.16 & ~ \cr
ASCA/SIS0 & 0.5-10 & 1994 Oct 11-12 & 42 & 0.21 & ~ \cr
ASCA/SIS1 & 0.5-10 & 1994 Oct 11-12 & 41 & 0.16 & ~ \cr
BeppoSAX/LECS & 0.1-10 & 1996 Sep 27 & 3.7 & 0.02 & ~ \cr
BeppoSAX/MECS & 1.5-10 & 1996 Sep 27 & 12.5 & 0.12 & 6.8$\times$10$^{-12}$ \cr
~ & ~ & ~ & ~ & ~ & ~\cr
\hline
\hline
$^a$ after applying the selection criteria\cr 
$^b$ single--temperature plasma model
\end{tabular}
\end{table*}

X--ray emission from the polar system BL Hyi (H01319-68) was first detected 
by HEAO-1/A-2, and later on by Einstein (Agrawal et al. 1983),
EXOSAT (Beuermann \& Schwope 1989, hereafter B\&S89)
and ROSAT (Ramsay et al. 1996; Schwope et al. 1997).   
In particular, the HEAO-1 and 
EXOSAT observations showed that copious soft X-ray 
emission originates from the secondary pole in
states of high accretion rate, while the EXOSAT observations 
showed this pole to become increasingly less active at lower accretion rates. 
This suggests a picture in which the source switches from one-- 
to two--pole accretion in going from low to high states (B\&S89). 
Recent EUVE observations indicate that the soft X--ray emission is
best reproduced by a 17 eV blackbody (Szkody et al. 1997). 

Despite all the above observations, 
the hard X--ray spectrum of BL Hyi was still poorly known before
the launch of ASCA and BeppoSAX satellites. ASCA observed this source 
in October 1994, while BeppoSAX observed it in September
1996, in the context  of a Core Program devoted to study flux and spectral
variability of Polars on different timescales and on a wide energy range.

In this paper we present temporal and spectral analysis of both 
ASCA and BeppoSAX observations. Sec. 2 describes the observations and data
reduction, while in Sec. 3 and Sec. 4 
we present temporal and spectral results, respectively, which 
are then summarized and discussed in Sec. 5.

\section{Observations and data reduction}

\subsection{ASCA}

The ASCA satellite (Tanaka et al. 1994)
contains  two CCD's (SIS0 and SIS1) and two GSPC's (GIS2 and GIS3).
SIS's have a broader energy range (0.5--10 keV) and better spectral 
resolution ($\sim$150 eV, roughly constant with energy), while GIS's
have a narrower band (0.8--10 keV), poorer energy resolution (about 500 eV
at 6 keV) but higher sensitivity above a few keV. 

The ASCA observation, carried out on 1994 Oct 11/12, 
was retrieved from the public archive 
(preliminary results on this observation
were presented by Fujimoto \& Ishida 1995).
The log of the ASCA observation can be found in Table 1. 
The observation was performed in a mixture of 1-CCD and 2-CCD modes for the
SISs. After cleaning, and selecting BRIGHT2 mode for the SISs,
the effective exposure time reduced to about 42, 41, 43 and 43 ksec
for SIS0, SIS1, GIS2 and GIS3, respectively. Data analysis was performed 
using the {\sc xanadu} and {\sc ftools} packages available from NASA
Goddard HEASARC.
Images, light curves and spectra were created using {\sc xselect}.
Spectral fitting and temporal analysis were performed with {\sc xspec}
and {\sc xronos}, respectively. 

GIS and SIS spectra and light curves have been extracted
within a circular region of radius equal to 6' (GIS2 and GIS3),
4' (SIS0) and 3' (SIS1) centred on the source.

For the GISs, background levels have been estimated in an annulus centred
on the source with inner radius of 6' and outer radius of 12'. For the SISs,
the entire chip containing the source was used, after excluding a circle of
4.5' centred on the source. 

The adopted response matrices are those released
on March 1995 for the GIS, while for the SISs
they have been created with the routine {\sc sisrmg}
in the {\sc ftools3.6.0} environment, adopting the March 1997 CTI calibration 
file.

\subsection{BeppoSAX}

BeppoSAX (see Boella et al. 1997 for an overall description of the mission)
carries both Narrow and Wide field instruments. Two Wide
Field Cameras point to opposite directions to each other, and perpendicularly
to four co-aligned Narrow Field Instruments:
two imaging instruments, the Low Energy Concentrator
Spectrometer (LECS) and the Medium 
Energy Concentrator Spectrometer (MECS); and two
collimated high--energy instruments, the Phoswich Detector System 
(PDS) and the High Pressure Gas Scintillation Proportional Counter (HPGSPC). 

The MECS is actually 
composed of three units, working in the 1--10 keV energy range,
with a total effective area of $\sim$150 cm$^2$ at 6 keV. 
The energy resolution is  $\sim$8\% and the angular resolution is
$\sim$0.7~arcmin (FWHM) at 6~keV. The LECS has characteristics similar to
that of a single MECS unit in the overlapping band, but its energy band
extends down to 0.1 keV. 

BeppoSAX observed the source on 1996 Sep 27. The effective exposure time
for the MECS was 12.5 ksec, but unfortunately, due to instrument problems,
 only 3.7 ksec of LECS data were collected. The source was not detected by
the HPGSPC and the PDS. 
The log of the BeppoSAX observation  can be found in Table 1.

MECS and LECS spectra and light curves 
have been extracted from a 4' radius circle centred on the source;
the MECS event files have been added together
after equalization of MECS2 and MECS3 data to the
MECS1 energy--PI relationship.  (In the following, with the name MECS 
we denote the sum of the three units.) 
We selected data acquired with an angle higher than 5$^{\circ}$
with respect to the Earth limb.
The background subtraction was performed using
blank sky spectra extracted from the same region of the detector
field of view. The spectra were fitted using the calibration matrices
released on September 1997.

\section{Results. Light Curves}

\subsection{ASCA}

We first searched for periodicity in both GIS and SIS light curves.
A peak, consistent with the optical period (but with a large error), 
was found; we then used in the 
analysis the optical ephemeris of Schwope 
(private communication to Szkody et al.
1997), even if it may no longer be fully appropriate (Schwope, 
private communication):

\begin{equation}
T = 2444884.21926 + 0.0789150406 E ~~~~~(HJD)
\end{equation}

\begin{figure}
\epsfig{file=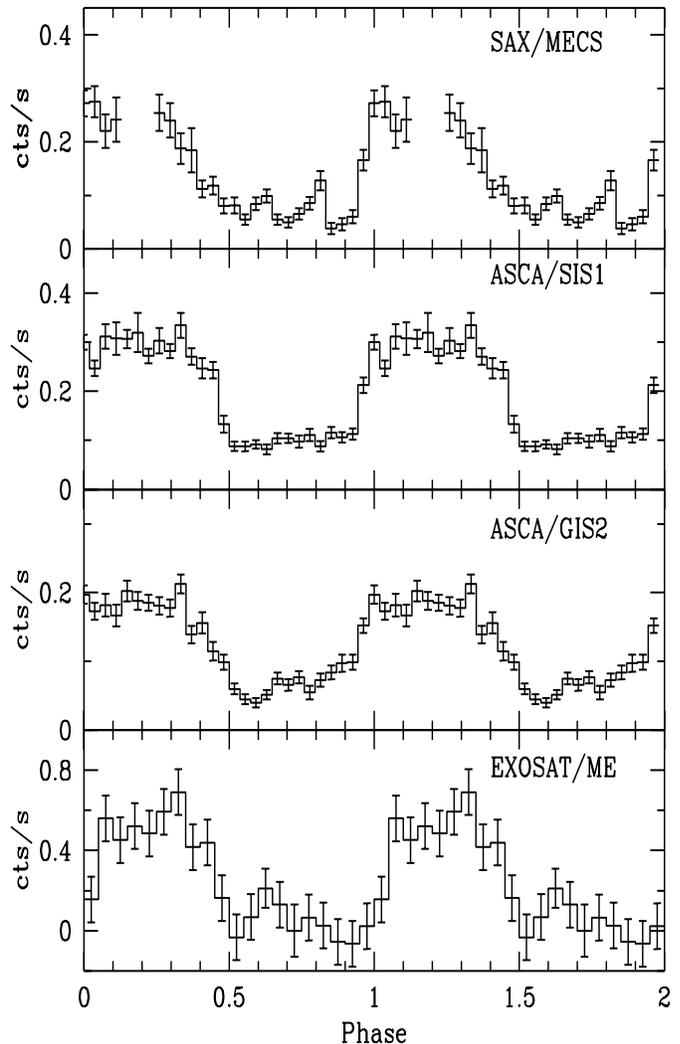, height=15.cm, width=11. cm, angle=0}
\caption{The light curve of the BeppoSAX/MECS (1.5--10 keV),
of the ASCA/GIS2 (0.8--10 keV) and SIS1 (0.5--10 keV)
and of the October 1985 observation
of the EXOSAT/ME (2--10 keV), from B\&S89. 
All curves are background subtracted.}
\label{lc}
\end{figure}

The corresponding folded light curves for the GIS2 and SIS1 are shown in 
Fig.~\ref{lc} (first and second panels from the top, respectively). The
two curves are rather similar, despite the SIS' broader and softer energy
band. Comparing the curves with that of B\&S89 (fourth panel), 
obtained with the EXOSAT--ME (i.e.
in a similar energy band) in October 1985, when the source was at a flux level
more than a factor 2 lower (see below), a rough agreement is found, 
but the maximum seems to start at a somewhat earlier phase in the ASCA curves.
Whether this difference is
due to an intrinsic difference in the accretion pattern or to the
ephemeris being no longer entirely appropriate is impossible to say until a new
improved ephemeris will be available. During the minimum, the source
is at a level much higher than the background, implying significant emission
from the second pole. 

A hardness ratio analysis was also performed for the SIS 
in order to search for spectral variations
along the period. No significant variations were observed, and in 
the spectral analysis (see section 4.1) we therefore used phase--averaged 
spectra. 

\subsection{BeppoSAX}

\begin{figure}
\epsfig{file=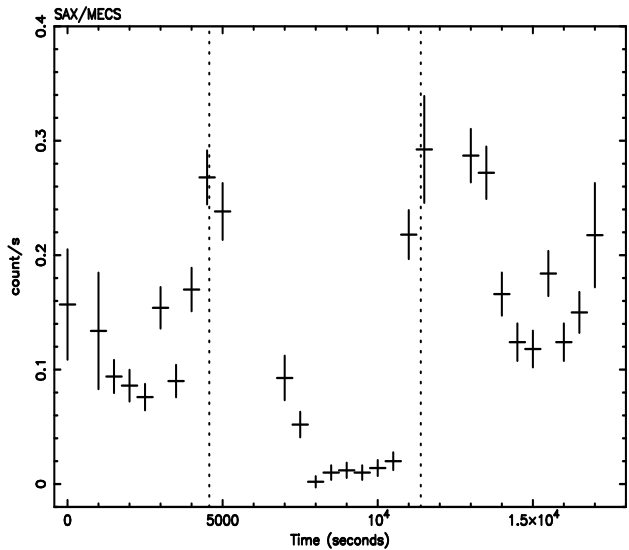, height=8.cm, width=9.5cm, angle=-90}
\caption{The light curve (background subtracted) of the BeppoSAX/MECS. 
The dotted, vertical lines mark phase zero.}
\label{mecs_lc}
\end{figure}

In Fig.~\ref{mecs_lc} and in the third panel of Fig.~\ref{lc} we show the MECS 
total and folded light curves, respectively. 
The folded light curve is incomplete due to the short
observing time. The onset is covered, however, and the curve
looks similar to the ASCA light curve. 

As in the ASCA light curve, there is significant emission outside the bright 
phase. It is also important to note that the minimum flux varies considerably
between individual orbits (Fig. 2). 
The BeppoSAX observation spanned about three cycles:
in the first and third the countrate is always well above the background, while 
in the second cycle the flux drops down to zero and then starts rising, but
remaining well below the corresponding level in the other
two cycles. The accretion on the second pole is therefore clearly variable, 
both from cycle to cycle and within each minimum. 

\section{Results. Spectra}

\subsection{ASCA}

As no spectral variations along the period are observed (see section 3.1),
we used phase--averaged spectra. 
To perform the spectral analysis, we summed together the two SISs and the two
GISs spectra, after having verified that the spectra of the single detectors
were consistent with each other. As said above (sec. 3.1), no significant
variations of the hardness ratio along the phase is observed (possibly
due to the limited statistics in the low phase) and we 
then used phase--averaged spectra
(which are, of course, dominated by the emission at the high phase).
The resulting spectra were grouped to have at least 20 counts per bin, in order 
to apply the $\chi^2$ statistics. As the
resulting SIS and GIS spectra were, in the overlapping band, consistent with 
each other, we fitted them simultaneously to improve the statistical quality
of the fits, keeping the two normalizations different to account for
possible remaining miscalibration between SIS and GIS. The results of the
fits are summarized in Table 2.

The fit with an optically thin isothermal plasma model (i.e. {\sc mekal}
model in {\sc xspec\_9.00}; model 1 in Table 2) is formally 
acceptable; it gives a 
temperature of $\sim$30 keV and a rather high value 
of the metal (essentially iron) abundance ($\sim$2.9 the solar one).
The 2--10 keV averaged flux in the GIS2 is 8$\times$10$^{-12}$ erg cm$^{-2}$
s$^{-1}$. 

Inspection of the residuals
(Fig.~\ref{ascasp}) indicates the presence of features, probably due
to absorption, at low energies (see below). Also, 
the model does not reproduce adequately the iron
line complex. Adding a gaussian line at $\sim$6.7 keV, i.e. from He--like iron,
a significant improvement is obtained, with a line 
EW of about 350 eV: this strongly suggests
that the real temperature is lower than estimated by the fit. Furthermore,
fitting the spectra in a restricted range with the upper energy fixed to 10 keV
and the lower  energy increasing from 1 to 6 keV, both the temperature
and the abundance monotonically 
decrease, going from 22 keV and 2.7 to 10.7 keV and 
0.7, respectively, again suggesting the presence of further components
in the spectrum as the reason for the high temperature and abundance found.

To investigate the origin of the observed low energy features in the residuals,
we first added a full cold absorber. The
quality of the fit increases significantly, and the plasma temperature 
reduces to $\sim$14 keV, a value more in line with the
common value for Polars (e.g. Ishida 1991; Done et al. 1995;
Beardmore et al. 1995; Ishida et al. 1997). The fit, however, is still
unsatisfactory because the value of the column
density, 8$\times$10$^{20}$ cm$^{-2}$, is largely in excess of 
that determined by EUVE and ROSAT ($\simlt 3\times$10$^{19}$ cm$^{-2}$: 
Szkody et al. 1997, Ramsay et al. 1996). 
We then fitted a partial covering model,
in analogy with other sources, in particular Intermediate Polars (Ishida 1991).
The fit (model 2) is, from a statistical point of view,
somewhat better than that with the full absorber; the absorbing column is about
3$\times$10$^{21}$ cm$^{-2}$, and the covering factor is $\sim$0.4. 
Fitting the spectra with more complex absorbers, like that discussed
by Done \& Magdziarz (1997), 
does not improve significantly the quality of the fit.

Even if the partial absorber is able to cure the low energy features in the
residuals and to significantly reduces the temperature, 
the abundance remains rather large (about 1.7). 
Such a high value has never been
observed in this type of sources and it will require a very specific 
evolution of the system. We then investigated
whether the high abundance could be an artifact of adopting 
a single--temperature model instead of a multi--temperature
plasma expected since a temperature gradient
along the emitting region is obtained from most models (e.g. Aizu 1973;
Wu et al. 1994; Woelk \& Beuermann 1996). The fit with the {\sc cemekl}
multi--temperature model, however, is worse than that with an isothermal
plasma and will not be considered further. 

We then included in the model the so--called  Compton
reflection component (Lightman \& White 1988;
Matt et al. 1991; van Teeseling et al. 1996) in order to account for 
reflection of the hard radiation by the white 
dwarf surface. The presence of this
component was suggested by
Done et al. (1995) and Beardmore et al. (1995) for EF Eri and AM Her, 
and it is now recognized to be an important ingredient of the X--ray 
spectrum of magnetic CVs. We adopted the model described in 
Done et al. (1995), assuming the matter
to be neutral (van Teeseling et al. 1996) and a mean inclination angle of
60$^{\circ}$. As the 
reflection continuum is always expected to be accompanied by a fluorescent
iron line at 6.4 keV (e.g. Matt et al. 1991), we added also this feature. 
As shown in Table 2 (model 3), an improvement in the fit is obtained,
and the abundance is now close to one. The best fit 
value for $R$ (i.e. $F(E)=F_0(E)[1 + R A(E)]$, where $F$ and $F_0$ denote
the total and primary flux, respectively, and $A(E)$ is the albedo of the
reflecting matter)
is rather high ($\sim$2.5), while a value about unity is 
expected for 2$\pi$ illuminated matter; this could be due to the fact
that a two--phase plane--parallel geometry is a too simple picture.
There is also a mismatch between the amount of reflection 
continuum and the EW of the 6.4 keV iron fluorescent
line: for the latter quantity, 
a value of about 90($\Delta\Omega/2\pi$) eV, obtained assuming a $\sim$10 keV
illuminating spectrum, solar abundances and a mean
observing angle of 60$^{\circ}$ is in fact expected,
to be compared with an observed 90\% upper limit of 70 eV. This 
 may suggest the presence in the spectrum of complexities too subtle to
be explored with the present data.

We also searched for Sulphur lines at $\sim$2.5 keV; 
contrary to the claim of Fujimoto 
\& Ishida (1995), based on the very same data, we found no
evidence for these lines.  Note that the SIS calibration
matrices have been quite improved since the time of their analysis,
which may explain the difference. 

In summary, there is clear evidence for complex absorption: a partial
covering model provides an adequate description of the data, even if more
complex absorbers cannot be excluded. The continuum is well described
by a single temperature plasma model (but, again, multi--temperature 
components cannot be rejected by the data), but in order to reduce the
metal abundance to ``normal" values, a reflection component is needed, 
similarly to the case of other Polars. 

\begin{table*}
\centering
\caption{
Best fit parameters for the ASCA (upper panel) ad BeppoSAX/MECS (lower panel)
observations. Errors correspond to
$\Delta\chi^2$=2.7, i.e. 90\% confidence level for one interesting parameter.
Model 1 is a single temperature plasma model with temperature $kT$ and
metal adundance (in units of the solar value) $A_Z$; 
models 2 and 3 include partial cold absorption with column density $N_H$ and
covering fraction C; model 3 includes also a cold reflection component with
relative normalization $R$ and a 6.4 keV iron fluorescent line.}
\label{GIS}
\vspace{0.05in}
\begin{tabular}{lccccccc}
\hline
\hline
~ & ~ & ~ & ~ & ~ & ~ & ~  \cr
\# & N$_{\rm H}$$^a$ (10$^{21}$ cm$^{-2}$) & C$^b$  & $kT$$^c$  (keV) 
& $A_Z$$^d$  & E.W.$^e$  (eV) & $R$$^f$  (10 keV) 
&  $\chi^2$/d.o.f. ($\chi^2_r$)   \cr
~ & ~ & ~ & ~ & ~ & ~ & ~ & ~  \cr
\noalign {\hrule}
~ & ~ & ~ & ~ & ~ & ~ & ~ & ~ \cr
%1 & ~ & ~ & 34.6$^{+6.6}_{-6.0}$ & 1 & ~ & ~ & 730/719 (1.02) \cr
1 & ~ & ~ & 31.0$^{+7.1}_{-6.5}$ & 2.86$^{+0.80}_{-0.67}$ & ~ & ~ & 704/718 
   (0.98) \cr
%3 & 0.77$^{+0.16}_{-0.16}$  & 1 & 14.1$^{+2.4}_{-1.8}$ & 2.00$^{+0.38}_{-0.33}$ 
%	& ~ & ~ & 631/717    (0.88) \cr
2 & 3.2$^{+2.4}_{-1.8}$ & 0.39$^{+0.20}_{-0.08}$ & 12.3$^{+1.4}_{-1.5}$ &
    1.76$^{+0.19}_{-0.20}$  & ~ & ~ & 624/716 (0.87) \cr
%5 & 2.9$^{+2.0}_{-1.7}$ & 0.40$^{+0.26}_{-0.08}$ & 12.5$^{+2.2}_{-1.6}$ &
%    1.70$^{+0.24}_{-0.30}$  & 72$^{+48}_{-53}$ & ~ & 618/715 (0.86) \cr
3 & 2.8$^{+3.4}_{-1.8}$ & 0.37$^{+0.32}_{-0.12}$ & 10.1$^{+1.9}_{-1.9}$ &
    0.96$^{+0.40}_{-0.31}$  & 20$^{+50}_{-20}$  & 2.5$^{+1.0}_{-1.1}$  & 609/714 (0.85) \cr
~ & ~ & ~ & ~ & ~ & ~ & ~ & ~ \cr
\hline
~ & ~ & ~ & ~ & ~ & ~ & ~ & ~ \cr
1 & ~ & ~ & 10.7$^{+3.4}_{-2.3}$ & 1.89$^{+0.92}_{-0.62}$ & ~ & ~ & 46/53 
   (0.87) \cr
~ & ~ & ~ & ~ & ~ & ~ & ~ & ~ \cr
\hline
\hline
\end{tabular}
~\par
$^a$ Column density of the absorber.\par
$^b$ Covering fraction of the absorber.\par
$^c$ Plasma temperature.\par
$^d$ Metal abundance in units of the cosmic value (Anders \& Grevesse 1989).\par
$^e$ Equivalent width of the 6.4 keV fluorescent iron line.\par
$^f$ Relative normalization of the reflection component (see text).
\end{table*}

\begin{figure}
\epsfig{file=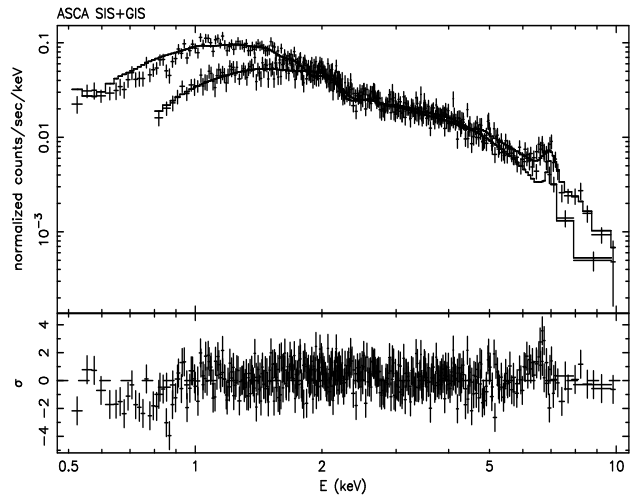, height=8.cm, width=9.5cm,  angle=-90}
\caption{The ASCA (SIS+GIS) spectrum fitted with an isothermal plasma model
(model 1 in Table 2.}
\label{ascasp}
\end{figure}

\subsection{BeppoSAX}

\begin{figure}
\epsfig{file=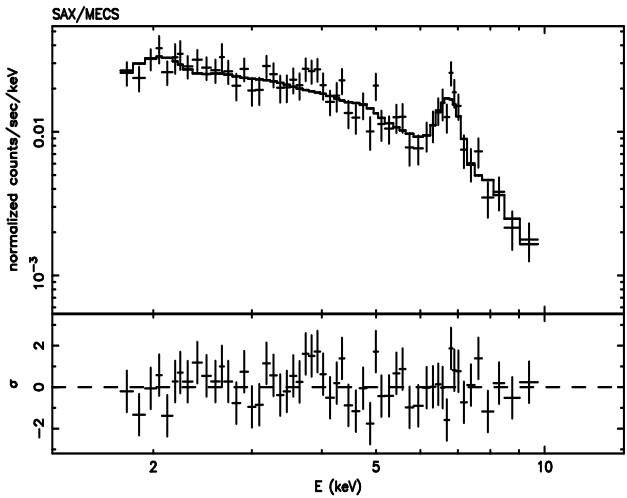, height=8.cm, width=9.5cm, angle=-90}
\caption{The BeppoSAX/MECS spectrum fitted with a thermal plasma model.}
\label{mecssp}
\end{figure}

The BeppoSAX/MECS spectrum of BL Hyi is shown in Fig.~\ref{mecssp}.   
The spectrum is well fitted by a single--temperature 
thermal plasma model with a temperature of 9.8$^{+3.0}_{-1.9}$ keV
and solar abundances
($\chi^2$= 52 with 54 d.o.f.). An improvement 
($\chi^2$= 46 with 53 d.o.f.) in the quality of the 
fit is obtained  if the metal abundance is left free to vary; the resulting
best fit values are: $kT$=10.7$^{+3.4}_{-2.3}$ keV, 
$A_Z$=1.89$^{+0.92}_{-0.62}$. The 2--10 keV averaged 
flux is about 6.8$\times$10$^{-12}$ erg cm$^{-2}$ s$^{-1}$. No significant
improvements in the fits are obtained with more complex models.

We then used the LECS to estimate the soft X--ray emission,  by adding a 
17 eV black--body component to the hard one, and assuming 
a column density of  3$\times$10$^{19}$ cm$^{-2}$ 
(Szkody et al. 1997). To be conservative, we put the black--body component
behind the cold partial screener (model 3 in table 2).
The combined LECS+MECS fit gives no black--body flux;
the 90\% upper limit to this component implies 
a bolometric soft luminosity less than about 10 
times that of the hard one. A much 
more stringent limit (i.e. $L_{\rm soft} < L_{\rm hard}$) can be put if
a 25 eV black--body temperature, as suggested by the ROSAT data (Ramsay et al. 
1996), in instead adopted.

\section{Discussion}

During the ASCA and BeppoSAX observations the source was significantly 
brighter than during the `intermediate state' of October 1985,
when it was observed by EXOSAT (B\&S89). Assuming a 10 keV thermal plasma
spectrum, the 2--10 keV phase--averaged 
flux, estimated with the {\sc pimms} package, 
 was in fact $\sim$3$\times$10$^{-12}$ erg cm$^{-2}$ s$^{-1}$
during that EXOSAT observation,  
while it was $\sim$2.3 and 2.7 times 
brighter during the BeppoSAX and ASCA observations, respectively. The ASCA
flux is in turn slightly lower than that 
measured by Ariel V ($\sim$10$^{-11}$  erg cm$^{-2}$ s$^{-1}$, Agrawal et al. 
1983). 

Comparing both ASCA and BeppoSAX folded light curves 
with the EXOSAT ones (and in particular with that of October 1985,
for which the statistics were the best), one important difference is apparent:
there is significant
emission from the second pole, while this emission during the EXOSAT 
observation was less intense, even if still present. As clearly
evident from the BeppoSAX light curve, 
the emission from the second pole is highly
variable, changing from cycle to cycle, and then suggesting non--stationary
accretion. 

The ASCA and BeppoSAX hard X--ray spectra are the first 
of sufficient quality to reveal spectral complexities. 
A simple thermal plasma model, even if formally acceptable, is unsatisfactory 
due to both the presence of features (mostly at the lowest energies) 
in the residuals and the high temperature and metal (mainly iron) 
abundance obtained. 
The inclusion of complex absorption cures the low energy
features. A partial covering model, with a column density of about 
3$\times$10$^{21}$ cm$^{-2}$ and a covering fraction of $\sim$0.4 provides
an acceptable description of the spectrum. 

It is conceivable that the absorber in front of 
the hard X-ray component is the
free-falling matter feeding the accretion spot. For example
in V834 Cen, this material
was identified by its Zeeman component shifted in wavelength as expected from
the free-fall velocity (Schwope \& Beuermann 1990). As the hard X-ray emitting
post-shock plasma is probably located in front of the soft X-ray emitting
sections of the photosphere, this absorbing material should affect the soft
component as well. Due to substantial pre-ionization, however, it may
be partially transparent to soft X-rays. On the other hand, the
fact that Zeeman absorption from this material was seen in V834 Cen indicates
that in that system neutral material is also present. Hence, the soft
component may not emerge unhindered and its spectral properties may carry
the signatures of the intervening absorber as well as those of the emitter. 
This has been taken into account when we have estimated
the intensity of the soft component from the BeppoSAX data. 
The upper limit so obtained 
is about ten times that of the hard luminosity, if a  black--body 
temperature of 17 eV (Szkody et al. 1997) is assumed; if, instead,
a temperature of 25 eV (Ramsay et al. 1995) is adopted the upper limit
is much more tight, reducing to about the hard luminosity. We also computed
the upper limit to the
black--body to hot plasma flux ratio in the 0.1--2.4 keV, to make comparison
with the theoretical expectation (see Fig.~8 in Beuermann 1997).
The limits are 5 and 1.3 for the adopted black--body temperatures, 
consistent with the comparatively hard X-ray
spectra encountered in low-field systems (Beuermann \& Schwope 1994). This is
especially true if the field strength of 12 MG measured by Schwope et al. (1995)
is characteristic of the main accreting pole. Even a field strength
of 23 MG reported by Ferrario et al. (1996) would still qualify BL Hyi as a
low-field system, characterized by a strong hard X-ray component. Only at still
higher field strengths is bremsstrahlung effectively suppressed by increasing
cyclotron cooling (Woelk \& Beuermann 1996, see also Beuermann 1997). 

Finally, the inclusion in the spectral model of the reflection 
component from the white dwarf surface provides a significant improvement
in the statistical quality of the fit and reduces the  abundance to 
roughly the solar value. This component has been already detected in a number
of Polars (e.g. Done et al. 1995; Beardmore et al. 1995; Ishida et al. 1997;
Cropper et al. 1997; Done \& Magdziarz 1997) and is now recognized as an
important ingredient of their X--ray spectrum. With all these components
included, the best fit 
post--shock temperature results to be about 10 keV 
(with about 20\% uncertainty) in both observations, consistent 
with the EXOSAT results (B\&S89) for the same source and with the values
usually found in Polars.

\begin{acknowledgements}
We acknowledge the BeppoSAX SDC team for providing pre--processed event files
and for their constant support in data reduction.
We thank A. Schwope for providing the EXOSAT light curves and for useful
discussions. 
GM acknowledges financial support from ASI, CD from a PPARC AF.
\end{acknowledgements}

\end{document}